\documentclass[epjST]{svjour}
\usepackage{graphicx}
\usepackage[version=3]{mhchem}				%
\usepackage{amsmath}
\usepackage{epstopdf}						%
\usepackage{epsfig}
\usepackage{subcaption}
        \graphicspath{{./_pic/}}              %
\usepackage[hidelinks]{hyperref}                   %

\begin{document}

\title{Destructive Impact of Molecular Noise on Nanoscale Electrochemical Oscillators}
\author{Filippo G. Cosi\inst{1} \and Katharina Krischer\inst{1} \fnmsep\thanks{\email{krischer@tum.de}}}
\institute{Physik-Department E19a, Technische Universit\"at M\"unchen, James-Franck-Strasse 1, D-85748 Garching, Germany}
\abstract{
We study the loss of coherence of electrochemical oscillations on meso- and nanosized electrodes with numeric simulations of the electrochemical master equation for a prototypical electrochemical oscillator, the hydrogen peroxide reduction on Pt electrodes in the presence of halides.
On nanoelectrodes, the electrode potential changes whenever a stochastic electron-transfer event takes place. Electrochemical reaction rate coefficients depend exponentially on the electrode potential and become thus fluctuating quantities as well. Therefore, also the transition rates between system states become time-dependent which constitutes a fundamental difference to purely chemical nanoscale oscillators. Three implications are demonstrated: (a) oscillations and steady states shift in phase space with decreasing system size, thereby also decreasing considerably the oscillating parameter regions; (b) the minimal number of molecules necessary to support correlated oscillations is more than 10 times as large as for nanoscale chemical oscillators; (c) the relation between correlation time and variance of the period of the oscillations predicted for chemical oscillators in the weak noise limit is only fulfilled in a very restricted parameter range for the electrochemical nano-oscillator.
} %

\maketitle

\section{Introduction}

Self-sustained oscillations in chemical systems are a prominent manifestation of self-organization under non-equilibrium conditions \cite{Nicolis:1995}. They occur in laboratory systems as well as in nature, the circadian clock governing pivotal functions of our body being just one prototype example \cite{Goldbeter2002}. The small volume of single cells, in which oscillations might take place, as well as nano-sized catalytic surfaces that support oscillatory surface reactions \cite{Kruse:2010,Barroo:2014} naturally lead to the question of the impact of molecular noise on the oscillatory behavior, or, in other words, of the minimum number of molecules necessary to support correlated oscillations. This question has been studied theoretically by Gaspard \cite{Gaspard2002}. For the weak noise limit, he derived an expression of the correlation time in terms of the extensivity parameter, the oscillation period, and the derivative of the period with respect to the Hamilton-Jacobi pseudoenergy. For a typical chemical oscillator this analysis predicts the lower limit of the number of molecules to be between 10 and 100. This result could be validated in recent experimental studies on the oscillatory $\mathrm{NO_2 + H_2}$ reaction on nano-sized Pt tips \cite{Barroo:2015}. Here, the product of the correlation time of the oscillations and the variance of the period of the oscillations was shown to be constant. 

Electrochemical oscillators differ from isothermal chemical oscillators because, in general, the electrostatic potential of the electrode is an oscillating quantity, in addition to the chemical variables \cite{Krischer:1999,Krischer:2002}. For nano- or mesoscale electrodes that are resistively coupled to some conducting support each stochastically occurring discrete electron transfer event changes the electrostatic potential of the electrode. At the nanoscale, the electrode potential is thus also a time-fluctuating quantity \cite{VGM1,VGM2} and so are the reaction rate coefficients, which depend exponentially on the electrode potential. Moreover, fluctuations of the electrode potential enhance each electrochemical reaction rate compared to the macroscopic limit. This peculiar behavior is closely connected to the potentiostatic control of electrochemical reactions, where the potential difference between the working electrode, i.e., the electrode of interest - in our case the nanoscale electrode - and a reference electrode is held constant. This operation mode imposes a constraint on the evolution of the electrode potential and may lead to non-normal distributions of the variables as well as shifts of the average steady state current densities compared to the macroscopic case \cite{VGM2}. The dependence of the rate constant of the electrochemical oscillations on a time-fluctuating quantity complicates theoretical studies on nanoscale electrochemical oscillators considerably. The dynamics of the nanoscale system is governed by an electrochemical master equation (EME) with time-dependent transition rates for the electrochemical reactions and an evolution equation for the electrode potential coupled to the one of the probability distribution of the chemical concentrations. The EME can be solved using an extended Gillespie algorithm \cite{VGM1,JansenMC}. 

In this paper, we investigate electrochemical oscillations on nanoscale electrodes, and compare them with chemical oscillations in small systems. We do so with simulations of a prototypical potentiostatic oscillator, the reduction of hydrogen peroxide on Pt electrodes in the presence of halides \cite{Mukouyama:2001}. This oscillator belongs to the large class of N-shaped negative differential resistance oscillators. In particular, we determine a critical electrode size below which correlated oscillations do not exist. Furthermore, we discuss the transition of uncorrelated to correlated behavior and test the validity of theoretical results for nano-scale chemical oscillators in the weak noise limit for electrochemical oscillators in regimes where the probability distribution of the period of the oscillations attains a Gaussian shape.

\section{Electrochemical master equation for the oscillatory reduction of $\mathrm{H_2O_2}$ on Pt}

\subsection{Electrochemical master equation}

An electrochemical reaction occurs on the surface of an electrode that is immersed in an electrolyte and involves the transfer of an electron from a species dissolved in the electrolyte to the electrode (oxidation reaction) or from the electrode to a species in the electrolyte (reduction reaction). These electron transfer reactions might be preceded, accompanied or followed by further elementary steps analogous to the ones occurring in a heterogeneous surface reaction, i.e. diffusion to or from the electrode, adsorption on the electrode surface, surface diffusion, reactions with other species and desorption from the electrode surface. Due to the electron transfer steps the rate of an electrochemical reaction depends on the electrostatic potential of the electrode. According to the Butler-Volmer Equation, the potential dependence of a reaction $\rho$ can be expressed through a potential dependent rate constant $k_{\rho}$:

\begin{align}
k_{\rho}(\phi) = \widetilde{k}_{\rho}^0 \mathrm{e}^{c_{\rho}(\phi- {E}^{0}_{\rho})} 
= k_{\rho}^0 \mathrm{e}^{c_{\rho}\phi}
\label{eq:rate_constant_phi_dep}
\end{align}

Here $\phi$ is the electrode potential (with respect to a given reference electrode), $\widetilde{k}_{\rho}^0$ the rate constant at equilibrium potential $E^{0}_{\rho}$. $c_{\rho}$ is given by:

\begin{align}
c_{\rho} = \frac{(\beta_{\rho} - \alpha) |n_{\rho}| F}{{R}T}
\label{eq:crho_general}
\end{align}

with $\alpha$ the transfer coefficient, $n$ the number of transferred electrons, $F$ the Faraday constant, ${R}$ the ideal gas constant and $T$ the absolute temperature. The value of $\beta_{\rho}$ changes with the form of the reaction $\rho$: $\beta_{\rho} = 0$ for reduction reactions and $\beta_{\rho} = 1$ for oxidation reactions. 
Note that for $n_{\rho} = 0$, it follows that $c_{\rho} = 0$ and the reaction is a purely chemical reaction. 

As long as the electrode potential $\phi$ does not change with time, e.g. due to a perfectly conducting connection of the nanoelectrode and the outer electric circuit,  $k_\rho$ is a constant, and the changes in the number of molecules of the chemical species involved in the reactions are goverend by the chemical master equation:

\begin{align}
\frac{\mathrm{d} P(\vec{N},t)}{\mathrm{d} t} = \sum_{\rho=1}^{r} \left [ W_{\rho}(\vec{N}- \nu_{\rho},k_{\rho})P(\vec{N}-\nu_{\rho},t) - 
W_{\rho}(\vec{N},k_{\rho})P(\vec{N},t) \right ]
\label{eq:ECME}
\end{align}

Here, $P(\vec{N},t)$ is the probability that the system is at time $t$ in state $\vec{N}=(N_1,N_2...)$ where $N_i$ is the number of molecules of species $i$. $W_{\rho}(\vec{N},k_{\rho})$ is the stochastic rate or transition probability of reaction $\rho$, which is expressed in terms of the macroscopic rate constant $k_{\rho}$. Each probabilistic reaction event $\rho$ changes the $N_i$s according to their stoichiometric coefficients $\nu_{\rho,i}$ which are equal to the difference of the number of products and reactants of species $i$ in reaction $\rho$. 

\begin{figure}[!h]
\centering
\includegraphics[width=0.55\textwidth]{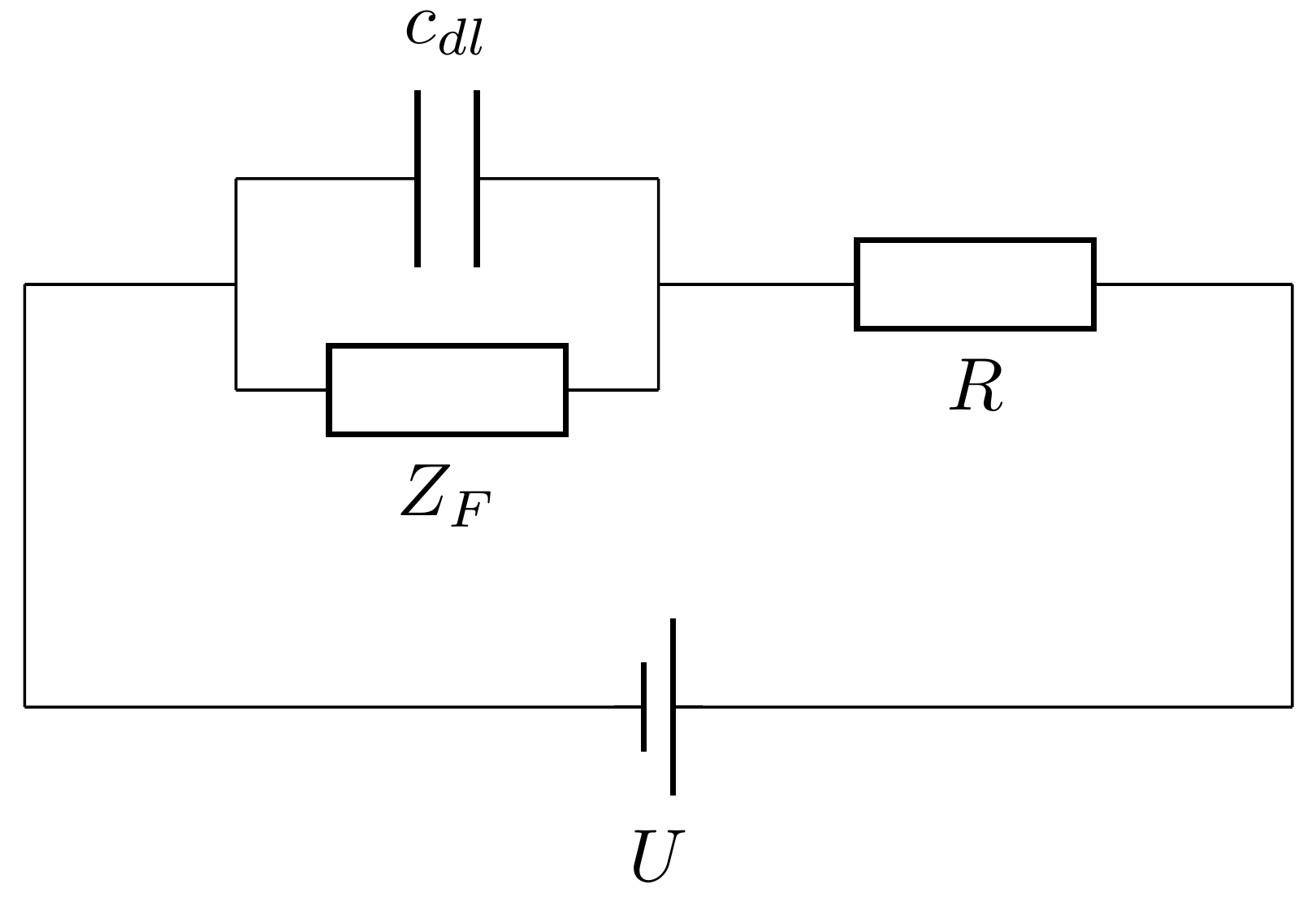}
\caption{Equivalent circuit for an electrochemical cell with double layer capacity per unit area $c_{dl}$, faradaic impedance $Z_{F}$, Ohmic resistance $R$ and applied voltage $U$.}
\label{fig:circuit}
\end{figure}

However, in general $\phi$ changes with time. This is in particular true for all electrochemical oscillators which rely on a direct or hidden negative differential resistance in the current $I-\phi$ curve where $\phi$ is the positive feedback variable \cite{Krischer:1999,Krischer:2002}. Figure \ref{fig:circuit} depicts the equivalent circuit of an electrochemical cell. In the deterministic limit the evolution equation for $\phi$ is obtained by applying Kirchhoff's circuit laws to the equivalent circuit:

\begin{align}
c_{dl} \frac{\mathrm{d} \phi}{\mathrm{d} t} = - i_{F} + \frac{U-\phi}{RA}
\label{eq:phi_ODE_macro}
\end{align}

where $c_{dl}$ represents the double layer capacitance per unit area, $i_{F}$ the faradaic current density and $A$ the electrode area. In a potentiostatically controlled experiment the externally fixed potential $U$ and the ohmic resistance $R$ are kept constant.

For meso- or nanoscale electrodes, a faradaic current flows each time an electron transfer reaction takes place. 
If we assume that after a time $\tau_{\rho}$ $n_{\rho}$ electrons are transferred to the electrode in a reaction $\rho$, the faradaic current density flowing due to this reaction event is given by $n_{\rho}e/A\tau_{\rho}$. Furthermore, since the time difference between two reaction events $\tau_{\rho}$ is very small, the change of the potential due to the potentiostatic control can be expressed as $\frac{U-\phi}{c_{dl}RA}\tau_{\rho}$. The macroscopic evolution equation for $\phi$ \eqref{eq:phi_ODE_macro} thus transforms to the following time-discrete equation for nanoscale electrodes:

\begin{align}
\phi_{j+1} = \phi_{j} - \frac{n_{\rho}\mathrm{e}}{c_{dl}A} + \frac{U-\phi}{c_{dl}RA}\tau_{\rho}
\label{eq:phi_evolution_micro}
\end{align}

The time dependence of $\phi$ renders $k_{\rho}$ time-dependent, which we will call from now on rate coefficient. Furthermore, due to the potentiostatic control, the potential changes also in between two reaction events, impeding the use of the classical Gillespie algorithm \cite{Gillespie:1976,Gillespie:1977} for solving the master equation \eqref{eq:ECME}. In particular, the formula for the waiting time between two reaction steps of the classical algorithms cannot be employed here. Instead, a generalized approach for the waiting times $\tau_{\rho}$ for each reaction $\rho$ to occur has to be employed \cite{VGM1,JansenMC}. 

\begin{align}
\tau_{\rho} = \frac{Rc_{dl}A }{c_{\rho}(U-\phi_{j})} \ln \left[ 1 + \frac{c_{\rho}(U-\phi_{j})}{W_{\rho}^0Rc_{dl}A \mathrm{e}^{c_{\rho}\phi_{j}}}\ln\left( \frac{1}{r_{\rho}} \right) \right]
\label{eq:taurho}
\end{align}
   
where $r_{\rho}$ is a random number between 0 and 1 that is associated with the probability of reaction $\rho$ to occur. The derivation of eq. (\ref{eq:taurho}) as well as a summary of the extended Gillespie algorithm employed in our simulations is given in the Appendix. Note that for $c_\rho=0$ $\tau_\rho$ takes on the value of the expression of the original Gillespie algorithm:

\begin{align}
\tau_{\rho} = \frac{1}{W_{\rho}^0}\ln \left ( \frac{1}{r_{\rho}} \right )
\label{eq:gillespie_waitingtime}
\end{align}

Furthermore, for convenience we express $W_{\rho}$ through a time-independent, chemical part, $W_{\rho}^0$, and a potential dependent term:

\begin{align}
W_{\rho} (\vec{N}) = W_{\rho}^{0} \mathrm{e}^{c_{\rho}\phi}
\label{eq:propensiety}
\end{align}

The dynamics of an electrochemical reaction network on a nanoscale electrode is thus determined by eqs. \eqref{eq:ECME} and \eqref{eq:phi_evolution_micro}, which are coupled by the potential-dependent transition rates eq. \eqref{eq:propensiety} and $\tau_{\rho}$ given by eq. \eqref{eq:taurho}.

\subsection{Oscillatory hydrogen peroxide reduction on Pt}

For our further analysis we will consider a particular oscillatory electrochemical reaction network, the hydrogen peroxide reduction in the presence of halide ions. This system was extensively studied by Nakato's group both experimentally and through mean field simulations \cite{Mukouyama:2001}. 
The oscillation mechanism involves the following reaction steps: %

\begin{align}
\label{eq:reaction1}
\ce{2Pt + H2O2} & \ce{->[k_1] 2(Pt-OH)} \\
\label{eq:reaction2}
\ce{(Pt-OH) + H+ + e-} & \ce{->[k_2] Pt + H2O} \\
\label{eq:reaction3}
\ce{2(Pt-OH)} & \ce{->[k_3] 2Pt +O2 + 2H+ + 2e-} \\
\label{eq:reaction45}
\ce{Pt + H+ + e-} & \ce{<=>[k_4][k_5] (Pt-H_{upd}) } \\
\label{eq:reaction67}
\ce{Pt + H+ + e-} & \ce{<=>[k_6][k_7] (Pt-H^{ot}) } \\
\label{eq:reaction8}
\ce{2(Pt-H^{ot})} & \ce{->[k_8] H2 +2Pt} \\
\label{eq:reaction910}
\ce{Pt + X- } & \ce{<=>[k_9][k_{10}] (Pt-X) + e- }
\end{align}

The notation $(\ce{Pt-M})$ represents the adsorption of a given molecule $\ce{M}$ on a platinum surface site, Pt. Reactions \eqref{eq:reaction2} and \eqref{eq:reaction8} are purely chemical, while all others involve an electron transfer step. The adsorption of upd (under potential deposition) hydrogen (forward reaction of eq. \eqref{eq:reaction45}) inhibits the adsorption (eq. \eqref{eq:reaction1}) and thus also the reduction (eq. \eqref{eq:reaction2}) of $\ce{H_2O_2}$, leading to a negative differential resistance (NDR) in the $I-\phi$ curve. Adsorption of halide ions (forward reaction of \eqref{eq:reaction910}) also inhibits $\ce{H_2O_2}$ adsorption but has an opposite potential dependence; $\ce{X-}$ desorbs when moving to negative potentials. Since the ad- and desorption dynamics of $\ce{X-}$ is much slower than the one of H-upd formation, the NDR is hidden for slow potential changes while it dominates the impedance at higher frequency. The oscillator is thus a classical 'hidden N-shaped negative differential resistance' (HN-NDR) oscillator \cite{Krischer:1999,Krischer:2002}. Note that in contrast to the original model in \cite{Mukouyama:2001} we neglect transport limitation of $\ce{H_2O_2}$ as well as the dependence of $k_1$ on $\ce{OH}$ and $\ce{X}$ coverage for the sake of simplicity. 
  
The transition rates $W_{\rho}^{0}$ corresponding to reactions \eqref{eq:reaction1}-\eqref{eq:reaction910} are compiled in table \ref{tab:propensities}, the rate constants and other parameters used in the simulations are given in tables \ref{tab:rateconst} and \ref{tab:redox_equil}, respectively, the initial values used in all simulations in table \ref{tab:initialval}. The extensivity parameter $\Omega$ is the absolute number of surface sites on the electrode surface:

\begin{align}
\Omega = {A\: n_s N_A}
\label{eq:area_electrochem}
\end{align}
	
Here, $N_A$ is the Avogadro constant and $n_s$ represents the active site density (given in $\frac{1}{mol\:cm^2}$). 

\begin{table}[!htb]
\centering
\begin{tabular}[c]{l c}
\hline
reaction & $W_{\rho}^0$ \\
\hline
\hline
\eqref{eq:reaction1}  & $\frac{k^0_{1}}{\Omega}C_{\ce{HO}}^b \left ( \Omega - N_{\ce{H}} - N_{\ce{HO}} - N_{\ce{X}} \right )$ \\
 & $ \left( \Omega - N_{\ce{H}} - N_{\ce{HO}} - N_{\ce{X}} -1 \right )$ \\
\eqref{eq:reaction2} & $k^0_2C_{\ce{H}}^b N_{\ce{HO}}$ \\
\eqref{eq:reaction3} & $\frac{k^0_3}{\Omega} N_{\ce{HO}}(N_{\ce{HO}} -1)$ \\
\eqref{eq:reaction45}$\rightarrow$ & $k^0_4 C_{\ce{H}}^b(\Omega - N_{\ce{H}} - N_{\ce{HO}} - N_{\ce{X}})$ \\
\eqref{eq:reaction45}$\leftarrow$ & $k^0_5 N_{\ce{H}}$ \\
\eqref{eq:reaction67}$\rightarrow$ & $k^0_6 C_{\ce{H}}^b(\Omega -N_{\ce{H}^{ot}})$ \\
\eqref{eq:reaction67}$\leftarrow$ & $k^0_7 N_{\ce{H}^{ot}}$ \\
\eqref{eq:reaction8} & $\frac{k^0_8}{\Omega}N_{\ce{H}^{ot}}(N_{\ce{H}^{ot}} -1)	$ \\
\eqref{eq:reaction910}$\rightarrow$ & $k^0_9 C_{\ce{X}}^s(\Omega - N_{\ce{H}} - N_{\ce{HO}} - N_{\ce{X}})$ \\
\eqref{eq:reaction910}$\leftarrow$ & $k^0_{10}N_{\ce{X}}$ \\
\hline
\end{tabular}
\caption{Preexponential factors of the transition rates.}
\label{tab:propensities}
\end{table}
\begin{table}[!htb]			%
\centering
\begin{subtable}[!htb]{0.49\textwidth}
\centering
\begin{tabular}{l  c l | l c}
\hline
$\widetilde{k}^0_{\rho}$ & Value &  &$c_{\rho}$ & Value $\left[\mathrm{V^{-1}}\right]$\\
\hline
\hline
$\widetilde{k}^0_1$ & $ 4.0 \cdot 10^{-2}$ &  $\mathrm{cm}\cdot\mathrm{s^{-1}}$ & $c_{1}$ & $0$\\
$\widetilde{k}^0_2$ & $ 1.0 \cdot 10^{-5} $ &  $\mathrm{cm }\cdot\mathrm{s^{-1}}$ & $c_{2}$ & $-19.33$ \\
$\widetilde{k}^0_3$ & $ 1.0 \cdot 10^{-8}$ &  $ \mathrm{mol}\cdot{\mathrm{s^{-1}}\cdot \mathrm{cm^{-2}}}$ & $c_{3}$ & $19.33$ \\
$\widetilde{k}^0_4$ & $ 1.0 \cdot 10^{-2} $ &  $\mathrm{cm }\cdot\mathrm{s^{-1}}$ & $c_{4}$ & $-11.60$ \\
$\widetilde{k}^0_5$ & $ 1.0 \cdot 10^{-5} $ &  $\mathrm{mol }\cdot{\mathrm{s^{-1}}\cdot \mathrm{cm^{-2}}}$ &  $c_{5}$& $11.60$ \\
$\widetilde{k}^0_6$ & $ 5.0 \cdot 10^{-3}$ &  $ \mathrm{cm }\cdot\mathrm{s^{-1}}$ & $c_{6}$ & $-19.33$ \\
$\widetilde{k}^0_7$ & $ 5.0 \cdot 10^{-6} $ &  $\mathrm{mol }\cdot{\mathrm{s^{-1}} \cdot\mathrm{cm^{-2}}}$ & $c_{7}$ & $-19.33$ \\
$\widetilde{k}^0_8$ & $ 5.0 \cdot 10^{-6} $ &  $\mathrm{mol }\cdot{\mathrm{s^{-1}}\cdot \mathrm{cm^{-2}}}$ & $c_{8}$ & $0$ \\
$\widetilde{k}^0_9$ & $ 5.0 \cdot 10^{-5} $ &  $\mathrm{cm }\cdot\mathrm{s^{-1}}$ & $c_{9}$ & $19.33$ \\
$\widetilde{k}^0_{10}$ & $ 5.0 \cdot 10^{-8}$ &  $ \mathrm{mol }\cdot{\mathrm{s^{-1}} \cdot\mathrm{cm^{-2}}}$ & $c_{10}$ & $-19.33$ \\
\hline
\end{tabular}
\caption{Time independent values of the reaction rate constants $\widetilde{k}_{\rho}^0$ and of the exponential coefficients $c_{\rho}$.}
\label{tab:rateconst}
\end{subtable}

\vspace*{0.5cm}

\begin{subtable}[!htb]{0.49\textwidth}
\centering
\begin{tabular}{l c | l c}
\hline
$E_{\rho}^0$ & Value & Parameter &  Value\\
\hline
\hline
$E_{2}^0 $  &  $0.8 \,\mathrm{V}$ &  $C_{\ce{HO}}^{b}$ & $0.7\cdot 10^{-3}\;\mathrm{mol}\cdot\mathrm{cm^{-3}}$\\
$E_{3}^0 $  &  $0.4 \,\mathrm{V}$ &  $C_{\ce{H}}^{b}$ & $0.3\cdot 10^{-3}
\;\mathrm{mol}\cdot\mathrm{cm^{-3}}$\\
$E_{4}^0 $  &  $-0.15 \,\mathrm{V}$ & $C_{\ce{X}}^{b}$ & $5\cdot 10^{-4}
\;\mathrm{mol}\cdot\mathrm{cm^{-3}}$\\
$E_{5}^0 $  &  $-0.15 \,\mathrm{V}$ & $N_s$ & $2.2\cdot 10^{-9} \;
\mathrm{mol}\cdot\mathrm{cm^{-2}}$ \\
$E_{6}^0 $  &  $-0.32 \,\mathrm{V}$ &$C_{dl}$ & $2.0 \cdot 10^{-5}
\;\mathrm{F}\cdot\mathrm{cm^{-2}}$\\
$E_{7}^0 $  &  $-0.32 \,\mathrm{V}$ & $T$ & $300\; \mathrm{K}$\\
$E_{9}^0 $  &  $-0.23 \,\mathrm{V}$ & $\alpha$ & 0.5\\
$E_{10}^0 $ & $-0.23 \,\mathrm{V}$  &\\
\hline
\end{tabular}
\caption{Values of the equilibrium redox potentials $E_{\rho}^0$ and further constants entering the transition rates $W_{\rho}$.}
\label{tab:redox_equil}
\end{subtable}
\vspace*{0.5cm}

\begin{subtable}[!htb]{0.49\textwidth}
\centering
\begin{tabular}{l  c}
\hline
Variable & $t=0$ \\
\hline
\hline
$\phi$  &  $-0.09$ $\mathrm{V}$ \\
$N_{\ce{H}}$  &  $0.1 \Omega$ 	\\
$N_{\ce{OH}}$  &  $0.1 \Omega$ 	\\
$N_{\ce{H}^{ot}}$  &  $0.4 \Omega$ 	\\
$N_{\ce{X}}$  &  $0.4 \Omega$ 	\\
\hline
\end{tabular}
\caption{Initial values of the dynamical variables used in the simulations of the $\ce{H2O2}$ reduction system.}
\label{tab:initialval}
\end{subtable}

\caption{Parameter values and initial conditions used in the simulations}
\end{table}

\begin{figure}[!h]
\centering
\includegraphics[width=0.65\textwidth]{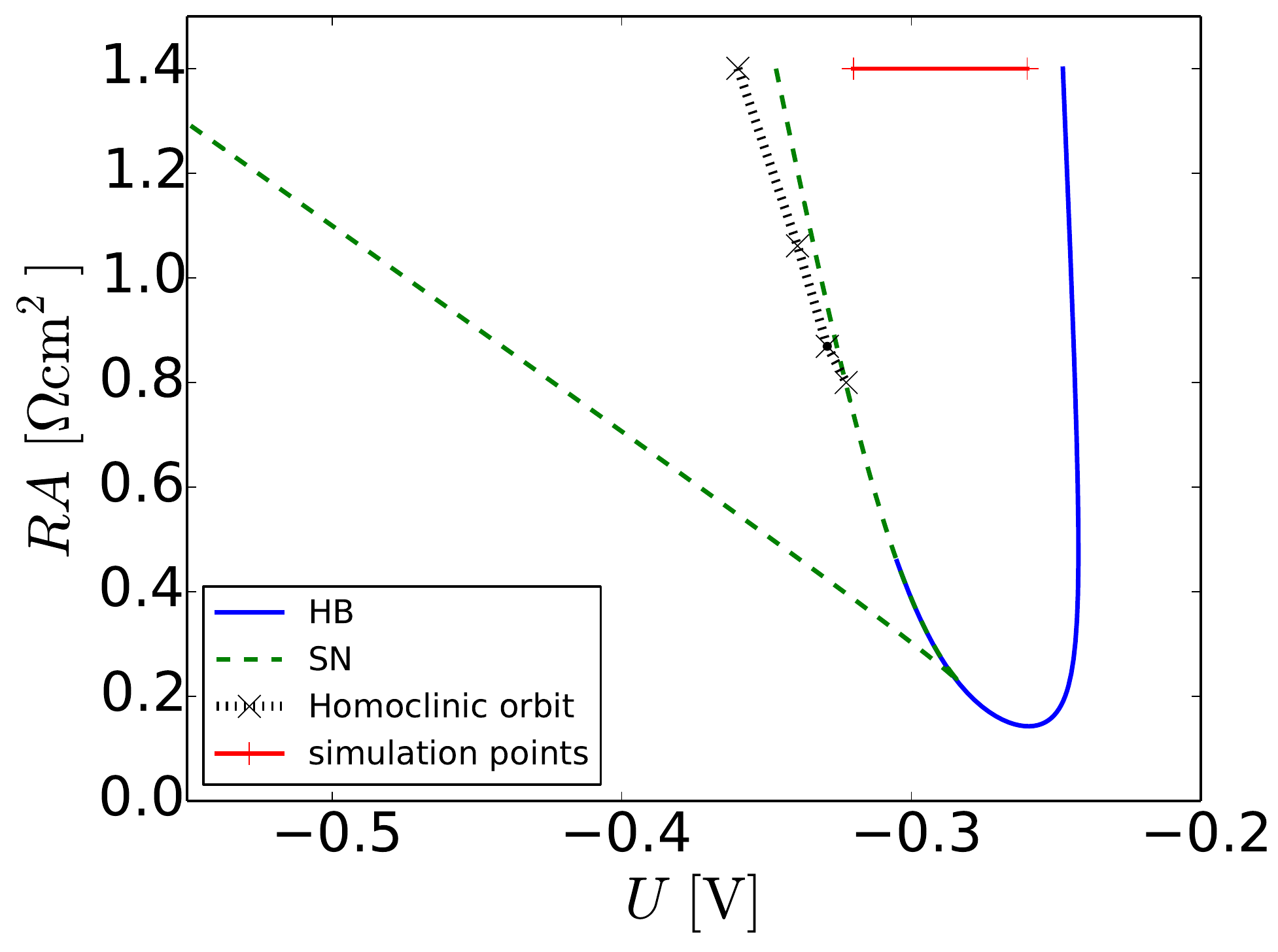}
\caption{Skeleton bifurcation diagram of the macroscopic system in the $R-U$ parameter plane. HB: Hopf bifurcation; SN: saddle node bifurcation; homoclinic orbits were calculated at the indicated crosses, the line connecting them being the approximte location of a saddle loop bifurcation.}
\label{fig:bifurcation_diagram}
\end{figure}

The bifurcation diagram of the macroscopic system is shown in fig. \ref{fig:bifurcation_diagram} in the resistance-applied voltage parameter plane. It was calculated from the deterministic rate equations for the evolution equations for $\phi$ (eq. \eqref{eq:phi_ODE_macro}) and the coverages of upd-H, OH, X, and H$^{ot}$ ('ot' standing for 'on top') using AUTO \cite{Doedel:1986}. Note that both $\phi$ and $U$ are given versus the saturated calomel electrode (SCE) throughout the paper. The evolution equations for the chemical species can be derived from the transition probabilities in table \ref{tab:propensities} taking the limit $\Omega \rightarrow \infty$. The solid and dashed lines in fig. \ref{fig:bifurcation_diagram} show the locations of Hopf and saddle node bifurcations, respectively. Thus, very roughly the system is oscillatory for parameter values between the Hopf line and the right saddle node line. The mesoscopic simulations we discuss in the following were done at 
$RA = 1.4 \: \mathrm{\Omega \cdot cm ^{2}}$ for different potential values along the red (dark gray) line in fig. \ref{fig:bifurcation_diagram}. At this resistance value the oscillatory region is bounded by a saddle loop bifurcation and a Hopf bifurcation at the low and high voltage borders, respectively. When traversing the oscillatory region from the high to the low voltage border the oscillations change from quasi-harmonic to spike-like relaxation oscillations.

\section{Simulations and Discussion}
\begin{figure}[!h]	
\centering
\includegraphics[width=0.65\textwidth]{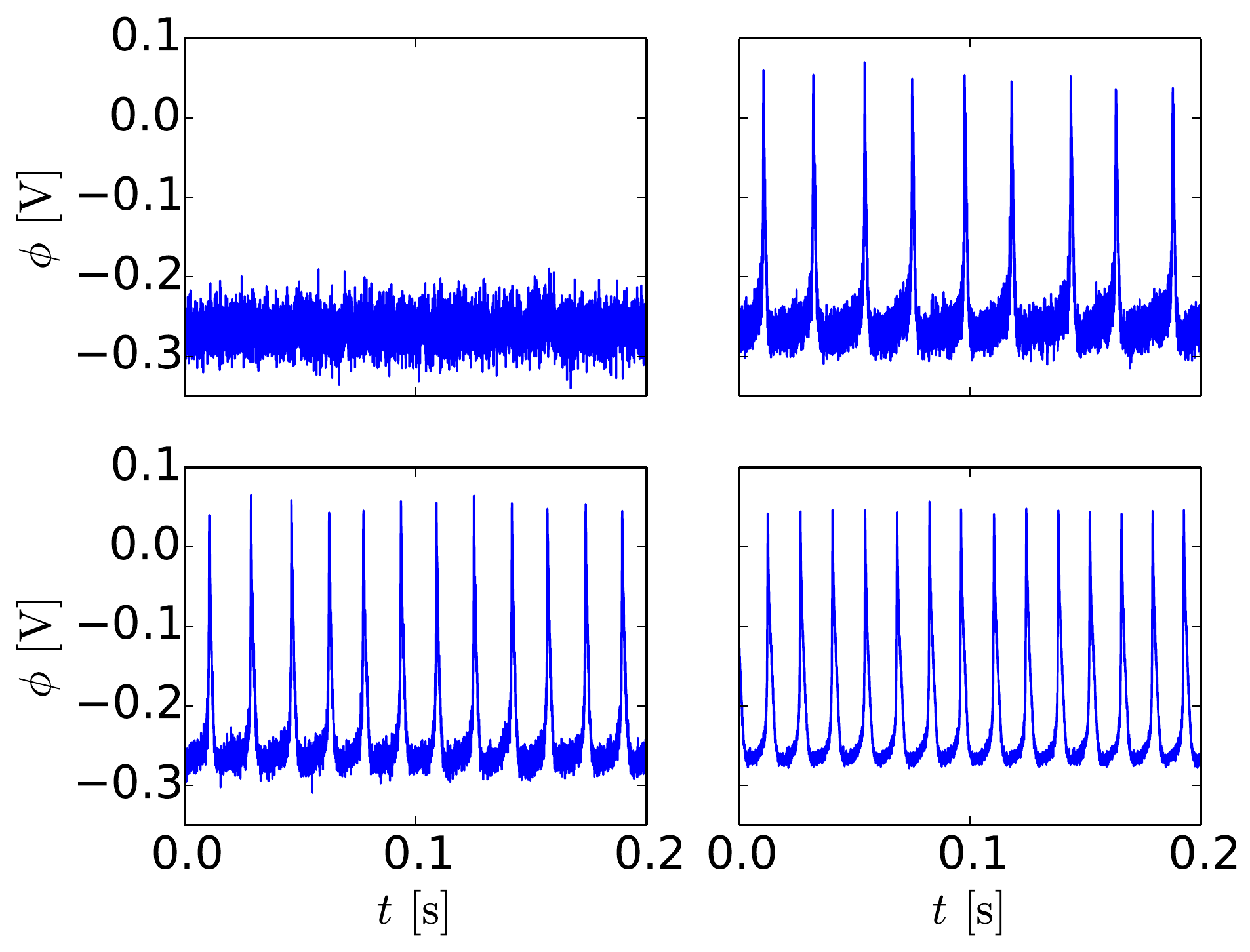}
\caption{Simulated time series of $\phi$ for values of $\Omega$ (from top left to bottom right): $1000$, $2000$, $4000$, $20000$ (corresponding to the macroscopic limit). Parameter values: $U = -0.30 \,\mathrm{V}$ and $RA = 1.4 \, \mathrm{\Omega}\mathrm{cm^2}$.}
\label{fig:oscill_series}
\end{figure}

\begin{figure}[!h]
\centering
\includegraphics[width=0.65\textwidth]{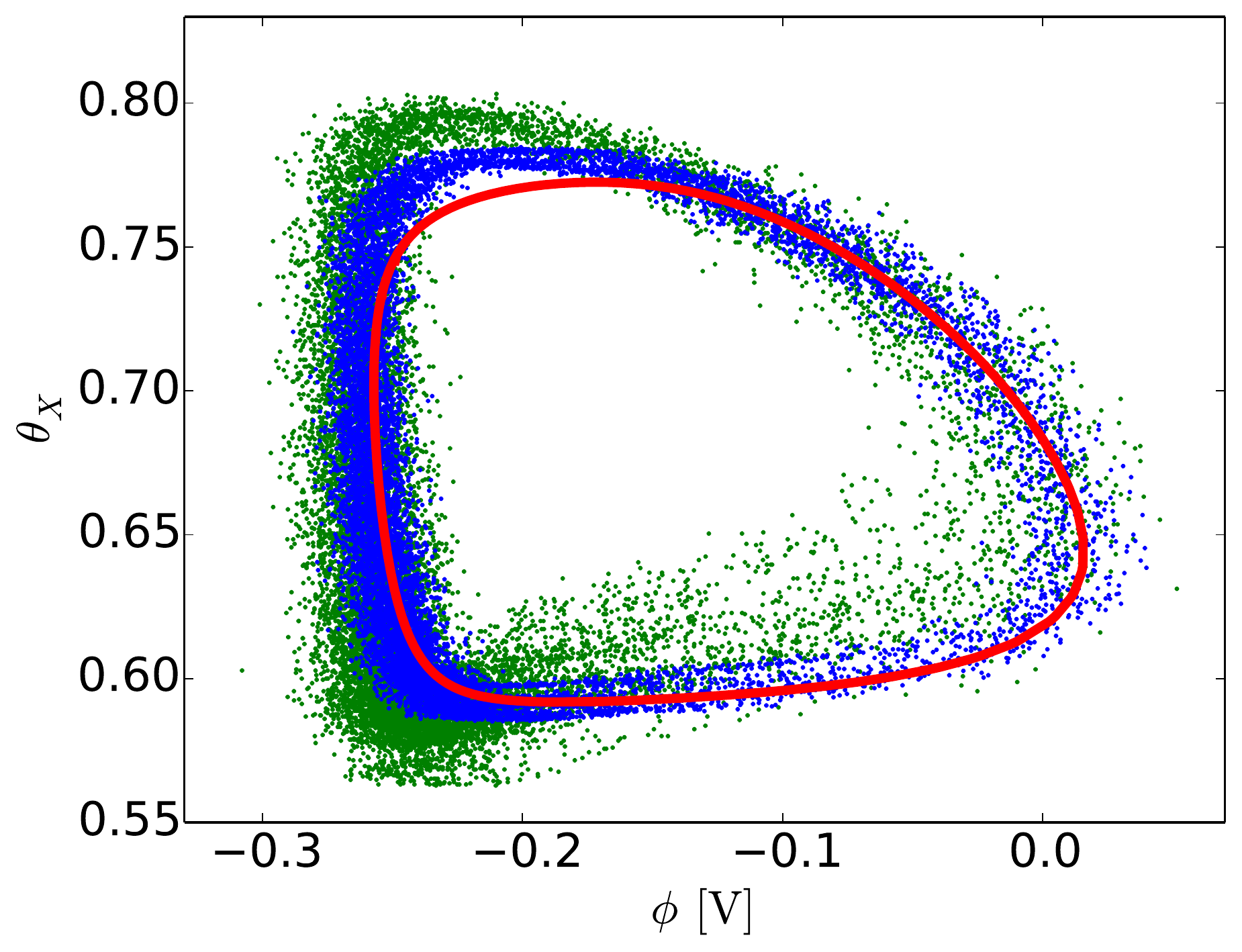}
\caption{Projections of trajectories on the $\theta_X$-$\phi$ plane for 
$\Omega = 2500$ (green respectively light gray) and $6000$ (blue respectively dark gray) limit cycle. The red (solid line) trajectory represents the macroscopic limit cycle. Parameter values: $U = -0.29\,\mathrm{V}$ and $RA = 1.4 \cdot 10^{-4}\,\mathrm{\Omega}\mathrm{cm^2}$.}
\label{fig:limit_cycles}
\end{figure}

In the following we discuss results obtained from stochastic simulations of the electrochemical master equation for the oscillatory $\ce{H2O2}$ reduction with the extended Gillespie algorithm. 
Time series of the electrode potential $\phi$ for four different system sizes $\Omega$ but otherwise identical parameters are depicted in Figure \ref{fig:oscill_series}. It is striking that the dynamics changes qualitatively with system size. From top left to bottom right the system size is increased, the lower right time series, obtained for $\Omega = 20000$, being very close to the macroscopic oscillations. For the smallest value of the extensivity parameter shown, $\Omega = 1000$ (top left), the system fluctuates around some mean value, however, without exhibiting any variations that would remind of an oscillatory excursion. Rather, it appears that the fluctuations are around a steady state. For intermediate system sizes, the system is still oscillatory, but the period of the oscillations is clearly larger than in the macroscopic limit. \\
A change of the average period with system size does not occur for purely chemical oscillators \cite{Gaspard2002}, which we also verified with the Brusselator model. We therefore attribute it to the peculiar property of electrochemical nanoscale systems that the reaction rate coefficients are fluctuating in time.
It was shown that this renders each individual electron transfer event faster on a nanoscopic than on a macroscopic electrode, whereby the slowest steps are enhanced strongest \cite{VGM1}. Hence, also the relative rates within the reaction network are changed, and we conjecture that decreasing the system size drives the system effectively through the saddle loop bifurcation which occurs at more negative voltages in the macroscopic system. Figure \ref{fig:limit_cycles} depicts a projection of three oscillatory trajectories obtained for different system sizes $\Omega$ in the $\phi-\theta_{\ce{X}}$ phase plane, $\theta_X$ being the coverage with halide ions. The continuous red line represents the macroscopic oscillation, the blue (dark gray) and green (light gray) lines are the trajectories in smaller systems. It is obvious that for the smaller systems the oscillations are deformed at low $\phi$ values towards higher $\theta_{\ce{X}}$ values, giving further evidence that for small system sizes the trajectories do not just fluctuate around the macroscopic limit cycle but occupy different regions in phase space. 

\begin{figure}[!h]
\centering
\includegraphics[width=0.65\textwidth]{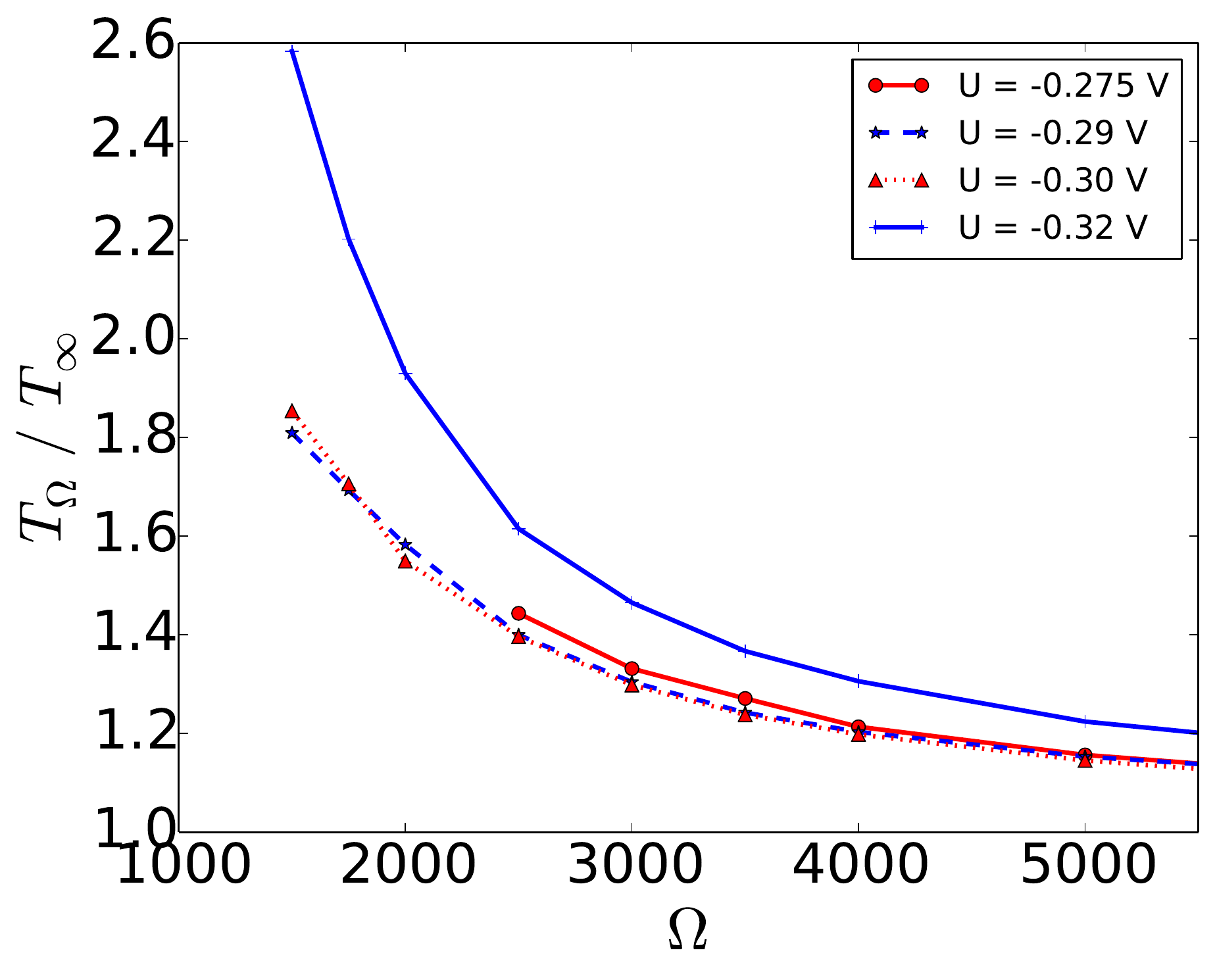}
\caption{Normalized oscillation period vs. extensivity parameter $\Omega$ for different values of $U$: $U = -0.275\,\mathrm{V}$ (vicinity of the Hopf bifurcation), $U = -0.29\,\mathrm{V}$, $U = -0.30\,\mathrm{V}$ and $U = -0.32\,\mathrm{V}$ (vicinity of the saddle-loop bifurcation).}
\label{fig:period_all}
\end{figure}

To quantify the trend of the mean oscillation period with $\Omega$, the fluctuating time series had been smoothed and the average return time of the resulting smooth oscillating time series (s.b.) determined. To this end, the return value was set above the noisy slow-passage part of the time series on the rising flank. More precisely, it was set to half the value between the maximum and minimum potential value of the oscillations. In some circumstances, especially for the more harmonic oscillations and small system sizes, this lead to multimodal distributions since here the amplitude fluctuations were so large that not all oscillations reached the set return value. In these cases, a smaller value was chosen by inspection such that the distributions became unimodal. The thus determined average return time, or mean period of the oscillations, $T_{\Omega}$, normalized to the period of the macroscopic limit cycle, $T_{\infty}$, is plotted vs. the system size $\Omega$ in fig. \ref{fig:period_all} for four different values of $U$. It can be clearly seen that the qualitative trend of an increase in mean period with decreasing system size is independent of the value of $U$. However, quantitatively, the increase is more pronounced close to the saddle loop bifurcation.

\begin{figure}[!h]	
\centering
\includegraphics[width=0.95\textwidth]{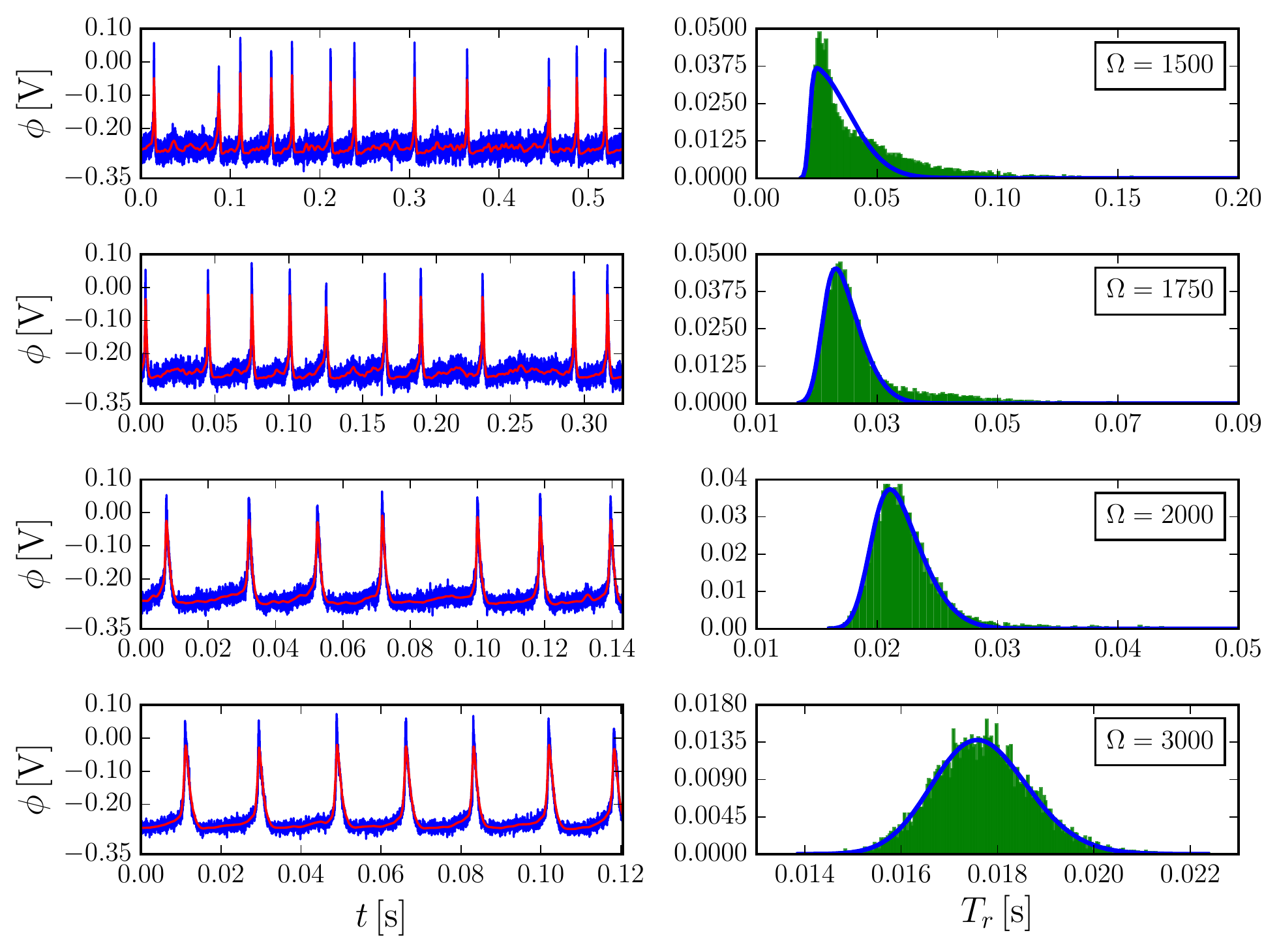}
\caption{Simulated time series (left plates) and probability distributions of the return time (right plates) for different values of $\Omega$ and $U = -0.30 \mathrm{V}$. Left plates: blue (dark gray) curves: full data, red (light gray) curves: smoothed data from which the return times were calculated; Right plates: blue (dark gray) curves: best fit of a skew-normal distribution. }
\label{fig:u30distributions}
\end{figure}

Next, let us investigate the robustness of the oscillations with respect to molecular noise. From Figs. \ref{fig:oscill_series} and \ref{fig:limit_cycles} it is obvious that the noise level in the time series, especially around the slow branch of the oscillations, increases with decreasing system size, as one would expect. First and foremost, molecular noise should manifest itself in a spread of the return times stemming from diffusion along the trajectories, i.e. so-called phase diffusion \cite{Kurrer:1991,Vance:1996,Gaspard2002}. Figure \ref{fig:u30distributions} displays four distributions of the return times obtained for different systems sizes and $U = -0.30 \mathrm{V}$ together with short parts of the respective time series. The distributions of the three smallest systems exhibit asymmetric tails towards higher periods and are thus clearly non-Gaussian. For the two smallest values of $\Omega$ the tails are so pronounced that it does not seem to be reasonable to assign an average or mean period to the time series. This interpretation is also supported by the corresponding autocorrelation functions (not shown) that fall off so quickly that the first maximum lies below $1/\mathrm{e}$. In addition, the time series for $\Omega = 1500$ reminds more of a noisy excitable system that fluctuates around a steady state where from time to time the fluctuations exceed a threshold such that an excursion in phase space is triggered. Owing to the nearby saddle loop bifurcation in the deterministic system and the excitable character of the global steady state beyond the saddle loop bifurcation together with the drift of the mean dynamics in phase space with system size, this scenario seems to be indeed conceivable. For increasingly larger systems, the distances between two successive oscillations are less spread, the mean and variance of the distribution being well defined from $\Omega = 3000$ on. 

For further evaluation, we fitted skew-normal distriubtions $\mathrm{SN}(x)$ to the data sets $x$, which can be represented by twice the product of a standard normal distribution $\varphi(x)$ and its relative cumulative distribution function $\Phi(\alpha x)$, the parameter $\alpha$ describing the asymmetry \cite{Azzalini:2013}.

\begin{align}
\mathrm{SN}(x) = 2\varphi (x) \Phi(\alpha x)
\end{align}

Considering the so called location $\chi$ and scale $\omega$, which correspond to the mean $\mu$ and the standard deviation $\sigma$, respectively, in the standard normal distribution, and using the fact that the cumulative distribution function depends linearly on the errorfunction $\mathrm{erf}$ of the dataset $x$, we can write the skew-normal distribution as follows:

\begin{align}
\mathrm{SN}(x) = \frac{1}{\omega\sqrt{2\pi}} {\exp}\left[{\frac{(x - \chi)^2}{2\omega ^2}}\right] \cdot \left [ 1 + \mathrm{erf}\left( \alpha \frac{x - \chi}{\omega} \right) \right ]
\label{eq:skewnormal}
\end{align}

Fitting the distribution of the return times with this expression allows us to determine the mean value $\mathrm{E}(x)$, the variance $\mathrm{var}(x)$ and the skewness $\zeta (x)$ from:

\begin{align*}
\mathrm{E}(x) = \chi + \omega \delta \sqrt{2/\pi} \\
\mathrm{var}(x) = \omega^{2} (1 - 2\delta^{2} / \pi ) \\
\zeta (x) = \frac{4 - \pi}{2} \frac{\mathrm{E}(x)^3}{\mathrm{var}(x)^{3/2}}
\end{align*}

where $\delta = \alpha / \sqrt{1 + \alpha^2}$.

The blue (dark gray) curves in the right column of fig. \ref{fig:u30distributions} represent the best fits of eq. \eqref{eq:skewnormal} to the distributions. The distribution for $\Omega = 3000$ has only a minor contribution of the skewness and can be considered Gaussian, the one for $\Omega = 2000$ can well be fitted by a skew-normal distribution, though with a non-negligible skewness of about 0.46, while in the other two cases the distributions are far from being describable by a skew-normal distribution. Thus, in these cases it is also not possible to determine a variance or standard deviation. Note also that whenever we talk about a mean return time or mean period of the nanoscale system that has an assymetric distribution, we refer to the maximum of the distributions, not to the average period. 

Figure \ref{fig:u30distributions} suggests that the loss of correlation of the oscillations for the spike-like oscillations at $U=-0.30 \;\mathrm{V}$ occurs through a lengthening of the interspike distances whereby the Gaussian distribution transforms first to a skew-normal distribution which is then further distorted towards longer distances until the spikes occur essentially uncorrelated. At still smaller system sizes the system fluctuates around a low potential value without exhibiting large amplitude spikes anymore (cf. fig. \ref{fig:oscill_series}).

\begin{figure}[!h]	
\centering
\includegraphics[width=0.95\textwidth]{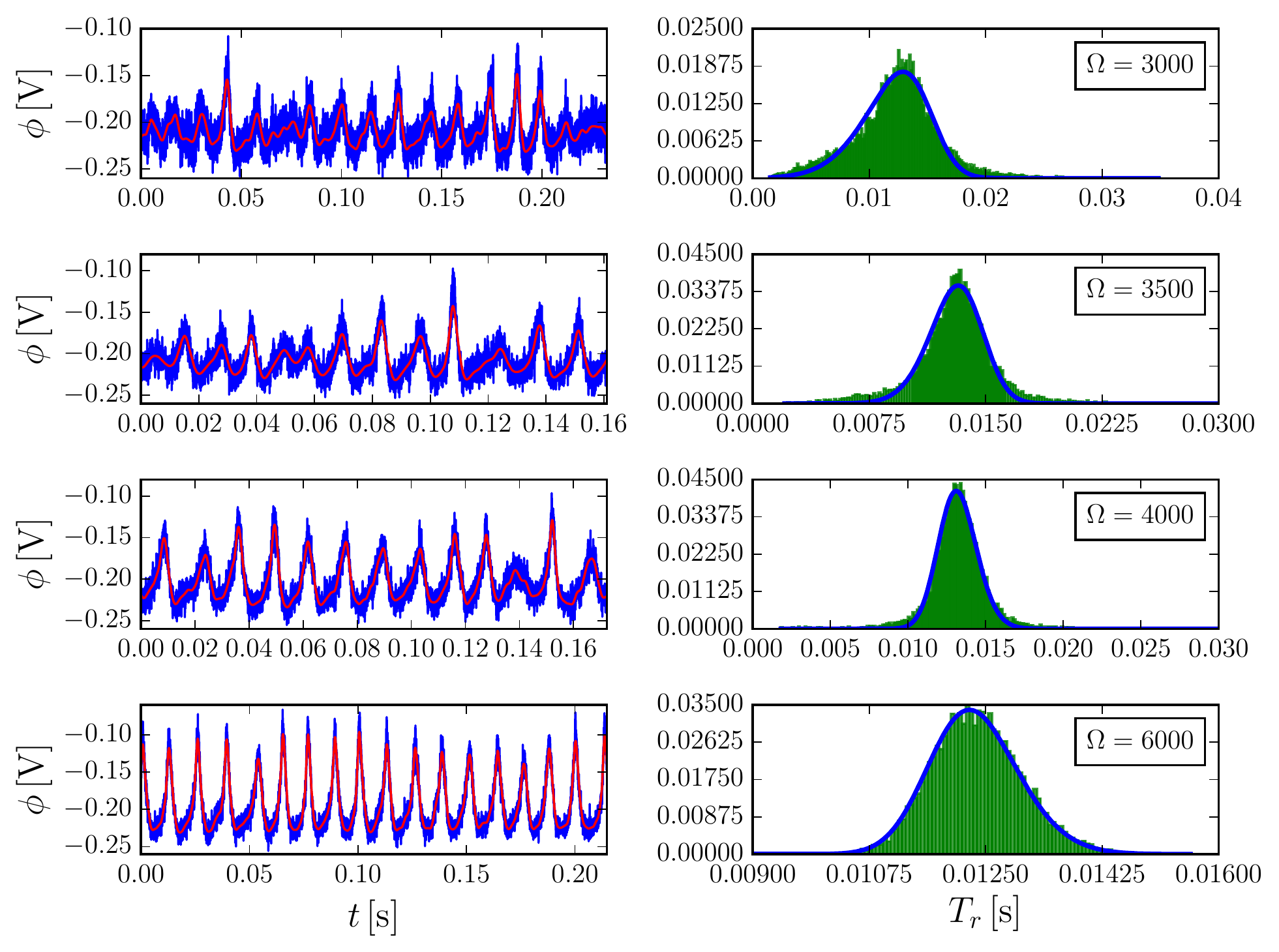}
\caption{Simulated time series (left plates) and probability distributions of the return time (right plates) for different values of $\Omega$ and $U = -0.26 \mathrm{V}$. Left plates: blue (dark gray) curves: full data, red (light gray) curves: smoothed data from which the return times were calculated; Right plates: blue (dark gray) curves: best fit of a skew-normal distribution.}
\label{fig:u26distributions}
\end{figure}

\begin{figure}[!h]	
\centering
\includegraphics[width=0.65\textwidth]{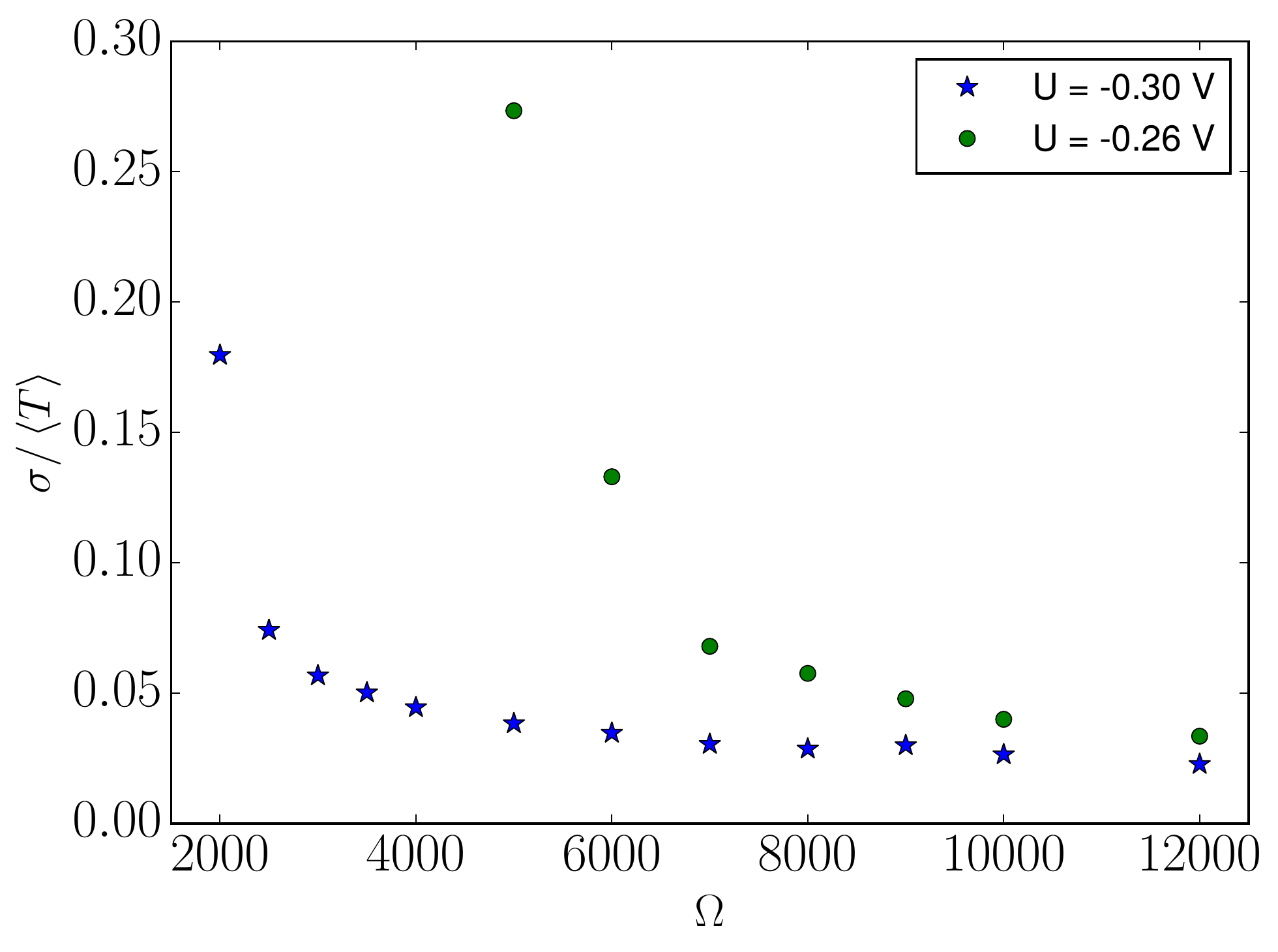}
\caption{Standard deviation of the return times normalized to the mean return time as a function of the extensivity parameter $\Omega$ for two different values of $U$ as indicated in the legend.}
\label{fig:systemsizes}
\end{figure}

This picture changes when the macroscopic base oscillations are more harmonic. Here, the probability distribution of the return times also progressively deviates from a Gaussian distribution when decreasing the extensivity parameter beyond a threshold value, but it stays approximately symmetric with respect to the maximum, exhibiting long-tails towards longer and shorter periods instead (fig. \ref{fig:u26distributions}). This difference also shows up in the time series. With decreasing system size they exhibit continuously higher noise levels and more strongly fluctuating amplitudes. Also the absolute system sizes at which coherent oscillatory behavior breaks down change considerably with the applied voltage, being much larger for the more harmonic oscillations at less negative potentials than for the spike-like relaxation oscillations at the negative potential border of the oscillations. This can be seen in fig. \ref{fig:systemsizes} where the standard deviations normalized to the mean period of those probability distribution functions (pdfs) that could be well fitted by a SN distribution are depicted for $U = -0.26\;\mathrm{V}$ and $ -0.30\;\mathrm{V}$. The relative standard deviation of the return times increases much faster with decreasing system size for $U = -0.26 \;\mathrm{V}$ than for $U = -0.30 \;\mathrm{V}$.

With a given lattice constant $a$, or average distance between two neighboring atoms, the extensivity parameter $\Omega$ can be related to an electrode size. Choosing $a = 0.27$ $\mathrm{nm}$ as a typical value, $\Omega_{min} = 2000$ as lower limit for correlated electrochemical oscillations at $U = -0.30 \mathrm{V}$  corresponds to an electrode area of about 12 x 12 nm$^2$ for flat electrodes or to spherical electrodes with a radius of about 3.5 nm. At $U = -0.26 \mathrm{V}$ the lower limit of $\Omega$ supporting regular oscillations was $4000$, which in turn translates to 17 x 17 nm$^2$ flat electrodes or spherical electrodes of 4.9 nm in radius, correspondingly. Compared to sizes reported in literature for heterogeneous surface reactions, these numbers are considerably larger. Also the number of molecules needed for a chemical clock, which was estimated to be between 10 and 100 \cite{Gaspard2002,Barroo:2015}, is close to 2 orders of magnitude smaller than for our electrochemical oscillator. For an electrochemical oscillator, the latter can be roughly estimated to be 1 to 2 times the value of $\Omega_{min}$, assuming that a typical oscillator has two to three types of adsorbed species with average coverages of 0.3 - 0.8. This implies that the noise level in electrochemical oscillators, where also the electrode potential fluctuates in time, is considerably larger than in chemical oscillators.  

Despite the strong dependence of the quality of a nonlinear oscillator on the characteristics of the limit cycle in the macroscopic system, theory predicts that in the low-noise limit where the chemical master equation reduces to a Fokker-Planck equation the product of the relative correlation time $\tau/\left\langle T\right\rangle$ and the relative variance of the  first return times $\sigma^2 / \left\langle T\right\rangle^2$ should be equal to $1 / 2\pi^2$ \cite{Barroo:2015}. For an electrochemical nanoscale system, the time-dependence of the rate coefficients $k$ hampers the derivation of a Fokker-Plank equation, and in fact makes it impossible to follow the same lines as for the a chemical system \cite{VanKampen}. Yet, it is possible to test this conjecture numerically. Therefore, we extracted the variance and mean period from the probability distributions of the period and calculated the autocorrelation function $C\left(t - t'\right)$ from the fluctuating time series. According to the theory $C\left(t - t'\right)$ should exhibit damped harmonic oscillations with an exponentially decreasing envelope, $\mathrm{e}^{-(t - t') / \tau}$, where the relative correlation time is the decay rate $\tau$ devided by the mean period $\left\langle T \right\rangle$. In the case of more harmonic oscillations and Gaussian pdfs, i.e. towards the high-potential border of the oscillatory region, the autocorrelation functions exhibited indeed exponentially damped, nearly harmonic oscillations. However, in case of the more spike-like oscillations towards lower potentials, the damped oscillations of the autocorrelation functions were neither harmonic in shape nor was their envelope exponentially decreasing. Rather, the decay of the maxima of the autocorrelation function was stronger than exponential, also in the cases of large $\Omega$ where the pdfs were Gaussian. In this case, we determined the delay time at which the envelope of the autocorrelation function attained the value $1 / \mathrm e$ by linear interpolation between the corresponding maxima of $C\left(t - t'\right)$ and used this delay time in the further evaluation. \\

\begin{figure}[!h]	
\centering
\includegraphics[width=0.65\textwidth]{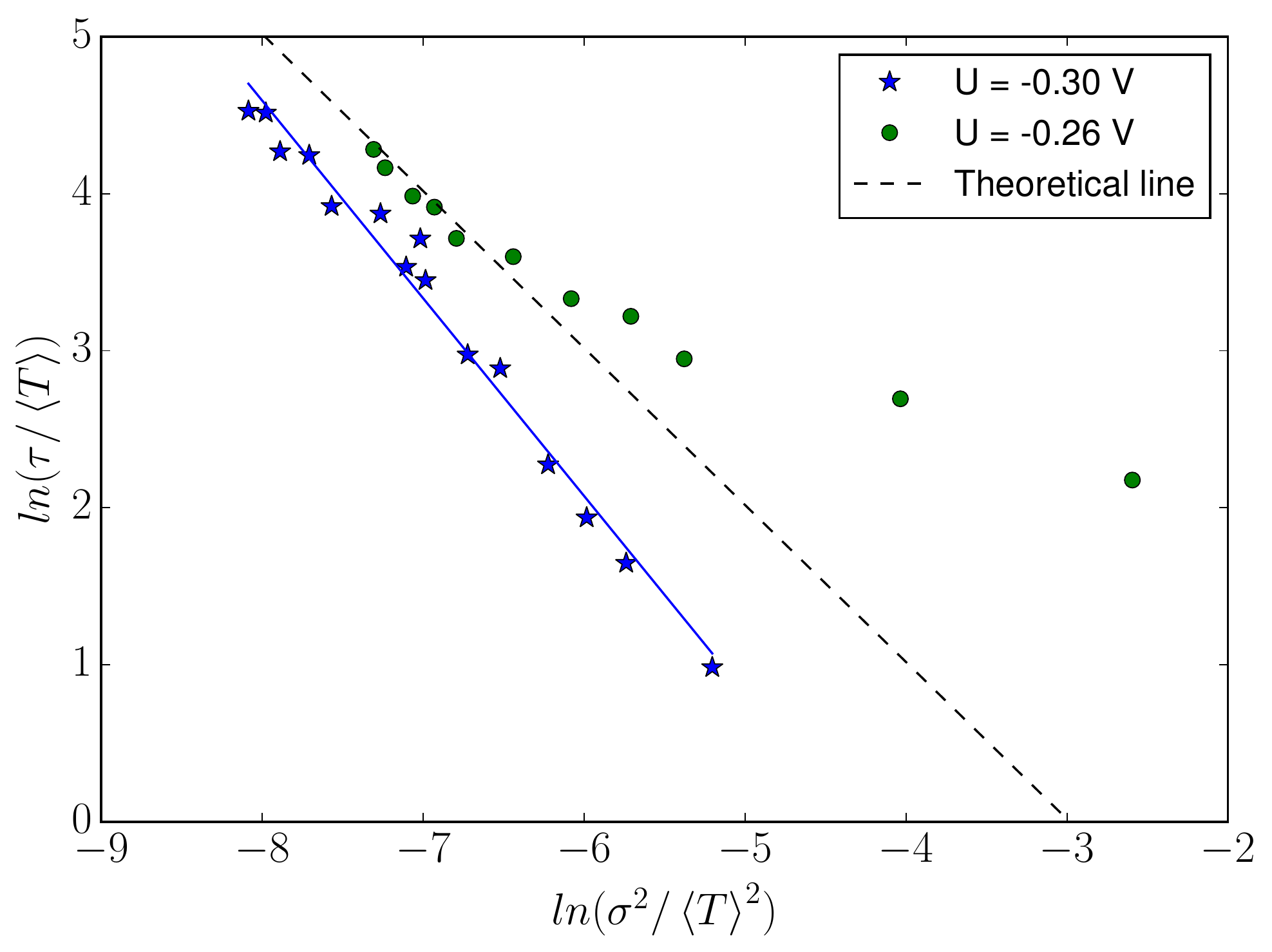}
\caption{Logarithm of the decay time of the autocorrelation function as a function of the logarithm of the relative variance of the return times for $U = -0.26 \mathrm{V}$ and $\Omega$ varying between 5000 and 20000 and $U = -0.30 \mathrm{V}$ and $\Omega$ varying between 2500 and 20000. Dashed line: theoretically predicted curve.}
\label{fig:tauvsvar}
\end{figure}

A double logarithmic plot of the relative decay time and the relative variance of the period of the oscillations is shown in fig. \ref{fig:tauvsvar} together with the theoretical predicted curve \cite{Barroo:2015} given by:

\begin{align}
\ln \frac{\tau}{\left \langle T \right \rangle} = - \ln \frac{\sigma ^2}{\left \langle T \right \rangle ^2} + \ln \frac{1}{2\pi^2}
\end{align}

 which has a slope of $-1$ and an intercept of $\ln(1 / 2\pi^2) \approx -2.98$. The data were
 obtained for two different $U$ values and system sizes that all have Gaussian distributions of the period. Clearly, the data of the two sets deviate significantly from the theoretical line. Furthermore, they separate in two groups corresponding to the $U$ values for which they were obtained. For $U = -0.30 \mathrm{V}$ the best linear fit gives a slope of $-1.26 \pm 0.05$ and an intercept of $-5.48 \pm 0.35$. This means that at a given relative variance of the period the relative correlation time is shorter than predicted, which probably reflects the faster than exponential decay of $C\left(t - t'\right)$ and the non-harmonic form of the oscillations. The situation is somewhat different for the more harmonic oscillations at $U = -0.26 \mathrm{V}$. Here, the data points obtained for the 5 largest system sizes between $\Omega = 20000$ and $12000$ lie on a line with a slope of $-1.03$ and an intercept of $-3.26$, very close to the values anticipated for the weak noise limit of chemical oscillators. However, for small system sizes the increase in relative variance is much stronger than predicted for the respective relative correlation times. Here, clearly the theory does not capture the observed behavior. We also determined the corresponding points for an intermediate value of $U$, where the data points showed a trend somewhat between the two displayed cases. Taken together, the data give evidence that the universal relationship between phase diffusion and correlation time of chemical oscillations is only applicable in very small parameter regions for electrochemical oscillations. At this point we can only speculate that the origin of this different behavior is a consequence of two properties of electrochemical systems, the stochastic time-dependence of the reaction rate coefficients, and the constraint on the dynamics introduced by the potentiostatic operation mode. For further insight more analytical work seems to be necessary. From this perspective, a break through seems to be linked to the success to derive a Fokker-Planck equation for electrochemical systems.

\section{Conclusions}
In conclusion, we demonstrated that electrochemical nanoscale oscillators feature some peculiar properties which render them less robust to molecular noise than chemical oscillators. The mean period of electrochemical oscillations on mesoscale and nanoscale electrodes depends on the system size, the mean trajectories of the oscillatory time series following different paths in phase space, and close to the macroscopic oscillation border the trajectories are driven out of the oscillatory region, thus confining oscillations to a smaller parameter range. Furthermore, the correlation time decreases strongly with decreasing system size, which might limit the smallness of nanoelectrodes in technological applications. The minimal system size and thus the lower bound to the number of chemical species necessary for time-correlated oscillations is one to two orders of magnitudes larger than for chemical systems, the minimal surface area of the electrode amounting to 100 to 500 nm$^2$. At sufficiently large system sizes the probability distributions of the first return periods are Gaussian, yet also in this region the universal relationship between correlation time and the variance of the period of the oscillations predicted for nanoscale chemical oscillations in the weak noise limit is not found in large parameter regions. Furthermore, for progressively smaller systems we found that the transition from coherent oscillations to noisy fluctuations lacking any defined mean period may occur through different scenarios. For spike-like relaxation oscillations the probability distribution of the next return map becomes more and more asymmetric towards larger periods before the correlation is lost completely, the oscillatory excursions of the smallest systems reminding of those in an excitable system. Thus, here the trajectories in phase space still follow rather defined paths, but at erratic times. For more smooth oscillations in the macroscopic limit, the probability distribution of the return times remains nearly symmetric but obtains long tails for system sizes below a threshold, the trajectories covering an increasingly larger region in phase space with decreasing $\Omega$ and thus the oscillatory cycles become less and less well defined. In the latter case, the system is also more prone to noise, coherent oscillations being lost already at larger system sizes than in the case of more relaxational oscillations. 

The decisive difference between the thus far studied chemical systems and the electrochemical system is that in the latter the electrode potential constitutes a further dynamical variable that renders the reaction rate coefficients to depend on time. Any chemical reaction which is not run under isothermal conditions will also involve changes of the temperature. Since all reaction rates depend exponentially on the temperature, a fluctuating temperature causes fluctuations in the reaction rate coefficient, just as the potential does in electrochemical systems. Dust particles in the higher atmosphere at which catalytic reactions occur might be an important realization of such nano- or mesoscale nonisothermal chemical systems. From this perspective, it would be interesting to study the impact of molecular noise in nonisothermal chemical nanoscale oscillators. They might exhibit some of the features we found for electrochemical systems, which would then be of more general nature. 

However, also the external control of electrochemcial systems enters into the dynamics in a nontrivial way. The potentiostatic operation mode introduces an external driving which renders all elemenatry reaction steps to be faster on a nanoelectrode than in the deterministic limit \cite{VGM2}. A deeper understanding of the peculiar features of the nano-oscillators discussed here requires further studies that allow the impact of the external driving to be separated from that of the time-dependent reaction rate coefficients.

\bibliographystyle{unsrt}

\begin{thebibliography}{}

\bibitem{Nicolis:1995}
G.~Nicolis, \textit{Introduction to Nonlinear Science}
(Cambridge Univ. Press, 1995)
%
%

%
%
%

\bibitem{Goldbeter2002}
D~Gonze, J~Halloy, and A~Goldbeter, Proc. Natl. Acad. Sci. U.S.A.
\textbf{{99}({2})} (2002) 673--678
%
%

\bibitem{Kruse:2010}
J.~S. McEwen, P.~Gaspard, Y.~De~Decker, C.~Barroo, T.V. de~Bocarme, and
  N.~Kruse, Langmuir
  \textbf{{26}({21})} (2010) 16381--16391
%
%
%

\bibitem{Barroo:2014}
C.~Barroo, Y.~De~Decker, T.V.~de~Bocarme, and N.~Kruse, J. Phys. Chem. C
\textbf{{118}({13})} (2014) 6839--6846
%
%
%
 %

\bibitem{Gaspard2002}
P.~Gaspard, J. Chem. Phys.
\textbf{117(19)} (2002)
%
%

\bibitem{Barroo:2015}
C.~Barroo, Y.~De~Decker, T.V.~de~Bocarme, and P.~Gaspard, J. Phys. Chem. Lett.
  \textbf{{6}({12})} (2015) 2189--2193
%
%
%
%

\bibitem{Krischer:1999}
K.~Krischer, \textit{Modern Aspects of Electrochemistry, Nr. 32}
(Kluwer Academic/ Plenum Publisher, 1999)
%
%

\bibitem{Krischer:2002}
K.~Krischer, \textit{Advances in Electrochemical Science and Engineering Vol. 8}
(Wiley-VCH Verlag GmbH \& Co. KGaA, 2002) 90--203
%
%
%

\bibitem{VGM1}
V.~Garc\'ia-Morales and K.~Krischer, Proc. Natl. Acad. Sci. U.S.A.
\textbf{107(10)} (2010) 4528--4532
%
%
 %

\bibitem{VGM2}
V.~Garc\'ia-Morales and K.~Krischer, J. Chem. Phys.
\textbf{134(24)}, (2011) 1--8
%
 %
%

\bibitem{JansenMC}
{A.P.J.~Jansen}, Comput. Phys. Commun.
\textbf{86(1–2)}, (1995) 1--12
%
%
%

\bibitem{Mukouyama:2001}
Y.~Mukouyama, S.~Nakanishi, T.~Chiba, K.~Murakoshi, and
  Y.~Nakato, J. Phys. Chem. B
  \textbf{105(30)}, (2001) 7246--7253
%
 %
 %
%

\bibitem{Gillespie:1976}
D.T.~Gillespie, J. Contemp. Phys.
\textbf{22}, (1976) 403--434
%
%
%

\bibitem{Gillespie:1977}
D.T.~Gillespie, J. Phys. Chem.
\textbf{81}, (1977) 2340--2361
%
%

\bibitem{Doedel:1986}
{E.~Doedel, Cong. Numer.}
\textbf{30}, (1981) 265--284
%
 %

\bibitem{Kurrer:1991}
C.~Kurrer and K.~Schulten, Physica D
\textbf{50(3)}, (1991) 311--320
%
%

\bibitem{Vance:1996}
W.~Vance and J.~Ross,
J. Chem. Phys. \textbf{{105}({2})}, (1996) 479--487
%
%
 %

\bibitem{Azzalini:2013}
A.~Azzalini and A.~Capitanio,
\textit{The Skew-Normal and Related Families.} (Cambridge Univ. Press, 2013)
%
%
 %

\bibitem{VanKampen}
N.G.~Van Kampen, 
\textit{Stochastic Processes in Physics and Chemistry.} (Elsevier, Amsterdam, The Netherlands, 3$^{rd}$ ed., 2007)
%
%
%
%
%
%
%
%
\end{thebibliography}

\newpage

\section*{Appendix}

\subsection*{Extended Gillespie Algorithm}

\subsubsection*{Derivation of Eq. \eqref{eq:taurho}}

The derivation given here follows closely the one given in the supplementary information of \cite{VGM1}.

The probability $p_{\rho}(\vec{N},t)$ that a reaction $\rho$ has not yet occurred after a time $t$ is given by the solution of the following differential equation

\begin{align}
\frac{\mathrm{d} p_{\rho}(\vec{N},t)}{\mathrm{d} t} = - W_{\rho}(\vec{N} + \nu_{\rho}|\vec{N})\:p_{\rho}(\vec{N},t)
\label{eq:noreaction}
\end{align}

According to eq. (\ref{eq:phi_evolution_micro}), in a small time interval $\Delta t$, in which no faradaic reaction occurs, the potential changes by

\begin{align}
\phi_{dl}(t + \Delta t) - \phi_{dl}(t) = \frac{U - \phi_{dl}(t)}{RAc_{dl}}\Delta t
\label{eq:phi_t_plus_deltat_without_n}
\end{align}

The probability that in the time interval $\tau_{\rho}$ reaction $\rho$ has not occurred can thus be expressed as

\begin{align}
P_{\rho}(\tau_{\rho}) = \exp \left ( - \int_{t_n}^{t_n+\tau_{\rho}} W_{\rho}\left(\vec{N} + \nu_{\rho}|\vec{N}\right )  \mathrm{d} t \right ) = \exp \left ( -W_{\rho}^0 \int_{t_n}^{t_n+\tau_{\rho}} \exp\left(c_{\rho} \phi_{dl}\right) \mathrm{d} t \right ) 
\label{eq:gillespie_probability_integration}
\end{align}

or, using eq. \ref{eq:phi_t_plus_deltat_without_n}

\begin{align}
P_{\rho}(\tau_{\rho}) = \exp \left [ -W_{\rho}^0 \int_{t_n}^{t_n+\tau_{\rho}} \exp\left ( c_{\rho}\phi_{dl}(t_{n}) + c_{\rho}\frac{U - \phi_{dl}(t)}{RAc_{dl}}(t-t_{n}) \right )  \mathrm{d} t \right ]
\label{eq:prob}
\end{align}

With the notation change to discrete potential steps $\phi_{dl}(t_n) \rightarrow \phi_{dl,j}$, integration of eq. \ref{eq:prob} yields:

\begin{align}
P_{\rho}(\tau_{\rho}) = \exp \left[ -\frac{W_{\rho}^0RAc_{dl} \exp\left(c_{\rho}\phi_{dl,j}\right)}{c_{\rho}(U-\phi_{dl,j})} \left( \exp\left( c_{\rho}\frac{U - \phi_{dl,j}}{RAc_{dl}}\tau_{\rho} \right) -1 \right)  \right]
\label{eq:extgillespie_prob}
\end{align}

Relating the random number $r_{\rho}$ to the probability $P_{\rho}(\tau_{\rho})$, i. e. setting $P_{\rho}(\tau_{\rho}) = r_{\rho}$, we can solve equation \eqref{eq:extgillespie_prob} for the waiting time $\tau_{\rho}$ and obtain:

\begin{align}
\tau_{\rho} = \frac{Rc_{dl}A }{c_{\rho}(U-\phi_{dl,j})} \ln \left[ 1 + \frac{c_{\rho}(U-\phi_{dl,j})}{W_{\rho}^0Rc_{dl}A \exp\left(
c_{\rho}\phi_{dl,j}\right)}\ln\left( \frac{1}{r_{\rho}} \right) \right]
\label{eq:extgillespie_waitingtime}
\end{align}

which is equal to eq. \eqref{eq:taurho}.

\subsubsection*{Gillespie Algorithm}

The electrochemical master equation was solved using Gillespie's generalized first reaction method. In detail the following steps were carried out:
\begin{enumerate}
	\item An input for $\phi$ and the species $N_i$ was chosen.
	\item The transition probabilities $W_{\rho}$ were calculated for the initial conditions.
  \item A random number $r_\rho$ was generated from the uniform distribution in the unit interval for each reaction $\rho$.
  \item Using eq. \eqref{eq:taurho}, a waiting time $\tau_{\rho}$ was calculated for each reaction $\rho$. 
  \item The equation with the smallest value of $\tau_{\rho}$ was chosen to advance. All $N_i$'s involved in the corresponding reaction are updated according to $N_i\rightarrow N_i + \nu^{i}_\rho$ and $\phi_j$ was updated according to eq. (\ref{eq:phi_evolution_micro}).
  \item The new propensities were calculated and steps 3 - 6 were repeated until the simulation was stopped, usually after several thousands of oscillation periods.
\end{enumerate}

\end{document}